\documentclass[aps,preprint,groupedaddress]{revtex4}  % for double-spaced preprint
\usepackage{amssymb}
\pdfoutput=1
\pdfoutput=1
\pdfoutput=1
\pdfoutput=1
\pdfoutput=1
\usepackage{amsmath}
\DeclareMathOperator{\sech}{sech}
\usepackage{amssymb}
\usepackage{graphicx}
\usepackage{subfigure}
\usepackage{color}
\usepackage{mathrsfs}
\usepackage[dvipsnames]{xcolor}

\usepackage[breaklinks=true,colorlinks=true]{hyperref}
\hypersetup{colorlinks=true,citecolor=blue,linkcolor=blue,urlcolor=blue}

\usepackage[utf8]{inputenc}

\RequirePackage{color}

\begin{document}
\immediate\write16{<<WARNING: LINEDRAW macros work with emTeX-dvivers
                    and other drivers supporting emTeX \special's
                    (dviscr, dvihplj, dvidot, dvips, dviwin, etc.) >>}

\title{Exact BPS double-kinks in generalized $\phi^4$, $\phi^6$ and sine-Gordon models}

\author{R. Casana$^{1,3}$}

\author{E. da Hora$^{2,3}$}

\author{F. C. Simas$^{1,3}$}

\affiliation{$^{1}$Departamento de Física, Universidade Federal do Maranhão, 65080-805, São Luís, Maranhão, Brazil.}
\affiliation{$^{2}$Coordenação do Curso de Bacharelado Interdisciplinar em Ciência e Tecnologia, Universidade Federal do Maranhão,  65080-805, São Luís, Maranhão, Brazil.}
\affiliation{$^3$Programa de Pós-graduação em Física, Universidade Federal do Maranhão, 65080-805, São Luís, Maranhão, Brazil.}

%%%%%%%%%%%%%%%%%%%%%%%%%%%%%%%%%%%%%%%%%%%%%%%%%

\begin{abstract}

We consider a $(1+1)$-dimensional theory with a single real scalar field $\phi$ whose kinematics is modified by a generalizing function $f(\phi)$. After briefly reviewing its Bogomol'nyi-Prasad-Sommerfield (BPS) structure, we focus on a particular $f(\phi)$ to obtain analytic BPS double-kink solutions in three different models governed by the $\phi^4$, $\phi^6$, and sine-Gordon superpotentials. In all cases, the resulting double-kinks approach the boundaries by following an exponential decay, with the generalizing function controlling its dependence on $x$ and mass. We also calculate the BPS bound explicitly and study how the double kinks behave near the origin. The energy distribution of the novel BPS states engenders symmetric two-lump profiles for the $\phi^4$ and sine-Gordon superpotentials. Whereas, for the $\phi^6$ superpotential, the BPS energy profiles form asymmetric two-lumps.

\end{abstract}

%%%%%%%%%%%%%%%%%%%%%%%%%%%%%%%%%%%%%%%%%%%%%%%%%

\maketitle

%%%%%%%%%%%%%%%%%%%%%%%%%%%%%%%%%%%%%%%%%%%%%%%%%
\section{Introduction}
\label{intro}
%%%%%%%%%%%%%%%%%%%%%%%%%%%%%%%%%%%%%%%%%%%%%%%%%

Configurations with nontrivial topology can be obtained as solutions to nonlinear field theories. Once the nonlinearity is commonly introduced via a symmetry-breaking potential, topological profiles are interpreted as the result of phase transitions \cite{masu}. These configurations have received attention due to their intricate physical aspects and possible applications.

The kink stands for the simplest topological object \cite{Enz,Fin,DHN}. Its canonical version emerges from a $(1+1)$-dimensional model with a single real scalar field only. Under special circumstances, it can be obtained also via a first-order differential equation that arises from the minimization of the energy according to the Bogomol'nyi-Prasad-Sommerfield (BPS) prescription \cite{prasom,bogo}. The resulting kink is then energetically stable once it saturates a well-defined energy bound.

The existence of multi-kink structures is a correlated issue that has gained attention in recent years. For instance, the creation of kink-antikink pairs in a two-dimensional $\phi^4$ theory was explored in Ref. \cite{Gri}, while an array composed by them was studied in Ref. \cite{Man}. In addition, a kink chain was considered in \cite{Vacha2}, while a kink-antikink lattice was investigated in Refs. \cite{Vacha1,Man2}. Even more recently, multi-kinks have also been obtained from enlarged two-field models \cite{blm,balbazmar,bazmarmen,jfd,marmen,bs,diosimas,hora,hora1}, from which they have been applied to explain experimental results in condensed matter physics \cite{poga,chdo}. Additional applications can also be found in Refs. \cite{Rev,Mat1,Mat2,Dun1,Dun2,Dun3}.

In particular, those multi-kink solutions that arise from higher-order scenarios are of special interest since these models may support long-range interactions, see Refs. \cite{c10,c11,c12,c13}. These configurations are expected to reveal relevant aspects under the influence of long-range forces that can affect the topological structure inherent to the interacting profiles, see Ref. \cite{c10}. The emergence of higher-order multi-kinks was considered in Refs. \cite{c14,c15}, while the scattering of multi-kink solutions was investigated in Ref. \cite{c16}.

In Ref. \cite{c17}, it was investigated a modification on the standard theory by incorporating an additional function in the Lagrangian density that depends solely on a single field. Additionally, in a recent work, Lima, Casana and Almeida have investigated novel BPS configurations in connection to a model with one scalar field only, see Ref. \cite{lca}. They have generalized the theory to include unconventional kinematics. In such a scenario, the authors have considered two different examples (i.e. the $\phi^4$ and $\phi^8$ superpotentials, separately), from which they have solved the BPS equation {\it{numerically}} for a non-polynomial generalizing function. As a result, they have obtained BPS double-kinks with a numerical profile.

Inspired by that investigation, we now go further by looking for {\it{exact analytic}} solutions that exhibit a double-kink behavior. With this aim in mind, we focus on a generalizing function that allows us to perform an analytical treatment of the first-order equation. To illustrate our construction, we consider the same generalizing function in three different models. Surprisingly, this same function holds well in all of them and led to double-kink profiles.

In order to present our results, this manuscript is organized as follows: in Sec. \ref{secII}, we introduce the model that contains one scalar field $\phi$ whose kinematics is extended through the inclusion of a generalizing function $f(\phi)$. We then implement the BPS framework to obtain the BPS bound and the self-dual equation whose solution saturates it. In Section \ref{secIII}, we particularize our investigation by considering a nontrivial $f(\phi)$ and, subsequently, analyzing three distinct superpotentials that generate the $\phi^4$, $\phi^6$, and sine-Gordon generalized models. In each case, we obtain analytical BPS solutions with a double-kink profile. We also show how $f$ influences the profile behavior near the boundaries ($x\rightarrow \pm\infty$) and close to the origin ($x \approx 0$). Furthermore, we examine the corresponding BPS energy distributions that present a two-lump format. One observes that the individual lumps that compose the profile may not be symmetric in relation to the origin, depending on the superpotential under investigation. To illustrate this aspect quantitatively, we compute the position and value of the maximum amplitude of all lumps. Finally, in Sec. \ref{secIV}, we present a summary and our perspectives regarding future contributions.

%%%%%%%%%%%%%%%%%%%%%%%%%%%%%%%%%%%%%%%%%%%%%%%%%
\section{The model and its BPS framework} \label{secII}
%%%%%%%%%%%%%%%%%%%%%%%%%%%%%%%%%%%%%%%%%%%%%%%%%

We consider a model with a single real scalar field $\phi(x,t)$. It is defined in a $(1+1)$-dimensional spacetime. Its kinetic term is enlarged to accommodate a generalizing function $f(\phi)$. As we demonstrate below, this function allows for the existence of analytical solutions with a double-kink profile. The corresponding Lagrange density is%
\begin{equation}
\mathcal{L}=\frac{1}{2}f\left( \phi \right) \partial _{\mu }\phi
\partial ^{\mu }\phi -V\left( \phi \right) \text{.} \label{ee1}
\end{equation}
Here, $f(\phi)$ is a nonnegative generalizing function, and $V\left( \phi \right)$ is the potential that describes the field self-interaction. It also defines the vacuum manifold of the model. We work in a Minkowski flat spacetime with a $\eta^{\mu \nu}=(+-)$ signature. The Greek index $\mu$ runs from $0$ to $1$. For the sake of simplicity, we consider all fields, coordinates and coupling constants dimensionless.

The model (\ref{ee1}) has its equation of motion (EoM) as
\begin{equation}
f \partial _{\mu } \partial ^{\mu }\phi + \frac{1}{2} f_{\phi} \partial _{\mu }\phi
\partial ^{\mu }\phi=-\frac{dV}{d \phi} \text{,} \label{ee2}
\end{equation}
where $f_{\phi}=df/d\phi$. In order to obtain topological solutions, it is necessary to consider a potential that allows for the spontaneous symmetry breaking mechanism. However, such $V(\phi)$ usually introduces nonlinear terms in (\ref{ee2}). As a consequence, the effective EoM can be quite hard to solve, even in the simplest $f=1$ scenario.

To circumvent this issue, we adopt an alternative approach. That is, we focus on those solutions that arise from a BPS framework. The construction of such a framework commonly requires the minimization of the energy related to (\ref{ee1}). With this in mind, we first observe that the time-independent energy density is simply given by $\varepsilon =-\mathcal{L}$. So, the corresponding total energy reads
\begin{equation}
E =\int \left[ \frac{f}{2}\left( \frac{d\phi }{dx}%
\right) ^{2}+V\right] dx\text{,} \label{ee3}
\end{equation}
where the integration runs over the entire $x$-axis.

Topological configurations must possess a localized energy distribution (or, equivalently, finite total energy). Therefore, the potential must satisfy
\begin{equation}
V(\phi \rightarrow \phi_\pm) \rightarrow 0\text{,} \label{cp}
\end{equation}
where $\phi_\pm=\phi(x\rightarrow\pm\infty)$ are fundamental states of the corresponding vacuum manifold. Also, $\phi'\sqrt{f}$ (prime denotes $d_x$ from now on) must vanish in the asymptotics. As we verify below, the exact double-kinks that we present in this manuscript satisfy all these conditions.

According to the BPS idea, we rewrite Eq. (\ref{ee3}) in the form
\begin{equation}
E =\int \left[ \frac{f}{2}\left( \frac{d\phi }{dx}\mp \sqrt{\frac{2V}{f}}%
\right) ^{2}\pm \sqrt{2fV}\frac{d\phi }{dx}\right] dx\text{.}\label{ee4}
\end{equation}

To proceed with the minimization, it is now necessary to choose the potential as
\begin{equation}
V\left( \phi \right) =\frac{W_{\phi }^{2}}{2f}\text{,} \label{ee5}
\end{equation}%
where $W_{\phi }=dW/d\phi$. Here, $W(\phi)$ represents the superpotential function. It is assumed to be regular for all $x$. In addition, $W_{\phi }\sqrt{f^{-1}}$ is supposed to vanish when $x \rightarrow \pm \infty$. This guarantees that $V(\phi)$ converges to a vacuum state, as predicted by (\ref{cp}).

Equation (\ref{ee5}) reveals that a nontrivial $f(\phi)$ changes the field self-interaction. So, it can be used to modify the vacuum structure of the effective BPS scenario. In this case, novel solutions are expected to emerge.

In view of the potential (\ref{ee5}), Eq. (\ref{ee4}) can be written as
\begin{equation}
E =\bar{E}+E_{BPS} \text{,} \label{b}
\end{equation}%
where we have defined
\begin{equation}
\bar{E}= \int \frac{f}{2}\left( \frac{d\phi }{dx}\mp \frac{W_{\phi }}{%
f}\right) ^{2}  dx \text{,}
\end{equation}%
and
\begin{equation}
E_{BPS}=\int \varepsilon _{BPS}dx=\pm \Delta W \text{,} \label{Ebps}
\end{equation}%
with $\Delta W=W\left( x\rightarrow +\infty \right) -W\left( x\rightarrow -\infty
\right)$. We have assumed that $W\left( x\rightarrow +\infty \right) \neq W\left( x\rightarrow -\infty
\right)$ (to be verified later below).

Equation (\ref{Ebps}) indicates that $E$ is submitted to a typical BPS inequality, i.e. $E =\bar{E}+E_{BPS} \geq \left\vert \Delta W\right\vert$. It means that the energy of the system is bounded from below. In this context, $E_{BPS}$ represents the value of such a bound, i.e. the minimum possible energy.

To saturate the lower bound, it is necessary to impose $\bar{E}=0$. As a consequence, one gets the first-order expression
\begin{equation}
\frac{d\phi }{%
dx}=\pm \frac{W_{\phi }}{f}\text{,}\label{EEbps}
\end{equation}
where the upper (lower) sign refers to the BPS kink (antikink).

The solutions that arise from Eq. (\ref{EEbps}) have an energy given by $E =E_{BPS}=\pm \Delta W$. Their energy density (i.e., the integrand of (\ref{Ebps})) reads
\begin{equation}
\varepsilon _{BPS}=\pm \frac{dW}{dx}\text{,}\label{EEx}
\end{equation}
which forces $W$ to be regular for all $x$, as previously noticed.

Based on the first-order expressions above, we study generalized scenarios that give rise to analytical BPS solutions with a double-kink profile. These scenarios are governed by the superpotentials that define the $\phi^4$, $\phi^6$ and sine-Gordon models, individually. They also contain a nontrivial expression for $f$. Surprisingly, the very same expression works well in all cases.

Before we proceed, it proves useful to use Eq. (\ref{EEbps}) to express the density (\ref{EEx}) as
\begin{equation}
\varepsilon _{BPS}=\frac{W_{\phi }^{2}}{f}\text{,}\label{edx}
\end{equation}
which connects a localized energy to a product $W_{\phi }\sqrt{f^{-1}}$ that vanishes in the asymptotics, see the discussion after Eq. (\ref{ee5}).

%%%%%%%%%%%%%%%%%%%%%%%%%%%%%%%%%%%%%%%%%%%%%%%%%
\section{Analytical BPS double-kinks} \label{secIII}
%%%%%%%%%%%%%%%%%%%%%%%%%%%%%%%%%%%%%%%%%%%%%%%%%

We now look for exact BPS solutions that engender a double-kink profile. Here, we follow an analytical prescription that can be applied to study novel configurations in various models. For the sake of illustration, we focus on the $\phi^4$, $\phi^6$ and sine-Gordon ones, separately.

First, we note that Eq. (\ref{EEbps}) can be written in the form
\begin{equation}
\frac{d\phi }{dy}=\pm W_{\phi }\text{,}\label{bebps}
\end{equation}%
where we have introduced the new coordinate $y$. The explicit relation between $y$ and $x$ must be obtained as the solution to
\begin{equation}
\frac{dy}{dx}=\frac{1}{f(\phi(x))}\text{.}\label{kk}
\end{equation}
So, it strongly depends on the expression for $f$ itself.

We now argue for a specific assumption (to be verified later below). As mentioned previously, we explore the $\phi^4$, $\phi^6$ and sine-Gordon superpotentials. In the canonical $f=1$ case, they lead to single-kink solutions $\phi_k(x)$ that can be inverted to provide $x(\phi_k)$. Here, we assume that such an inversion is still possible even for a nontrivial $f$. With this assumption in mind, in what follows, we specify $f$ as a function of $x$, instead of $\phi$.

Based on the arguments above, we choose the generalizing function as
\begin{equation}
f\left( x\right) = \left( \frac{2n+1}{x}\right)^{2n} \text{,} \label{ef1}
\end{equation}%
where $n \ge 0$ is an integer. For $n=0$, this function is equal to $1$. So, it recovers the standard scenario. In what follows, we show that Eq. (\ref{ef1}) generates the double-kink behavior of the resulting BPS solutions.

Note that, for $n\geq 1$, $f$ diverges as $x$ approaches the origin. In this case, for any regular superpotential $W$, Eq. (\ref{EEbps}) states that $\phi'(x)$ tends to zero near this region. As a consequence, a plateau is expected to appear. Note that this expectation does not depend on the particular expression for $W$. That is, the plateau is expected to occur in the $\phi^4$, $\phi^6$, and sine-Gordon cases. In fact, this is precisely what happens. We point out here that such a plateau around $x=0$ also exists in the purely numerical scenario; see Figs. 5 and 11 of Ref. \cite{lca}.

On the other hand, in the limits $x \rightarrow \pm \infty$, $f$ vanishes. In this case, $W_\phi$ is expected to vanish faster in order to ensure that $\phi$ approaches its vacuum states smoothly. This property is also present in the numerical results explored in \cite{lca} itself.

In view of (\ref{ef1}), Eq. (\ref{kk}) assumes the effective form
\begin{equation}
\frac{dy}{dx}=\left( \frac{x}{2n+1}\right) ^{2n}\text{,}
\end{equation}%
whose solution is%
\begin{equation}
y\left( x\right) =C_{n}x^{2n+1}\text{.}\label{yx}
\end{equation}%
Here, the integration constant was assumed to be $0$, for simplicity. Also, we have defined%
\begin{equation}
C_{n}=\frac{1}{\left( 2n+1\right) ^{2n+1}}\text{,}
\end{equation}%
for the sake of convenience.

The energy density (\ref{edx}) of the BPS solutions assumes the form
\begin{equation}
\varepsilon _{BPS}=\left( \frac{x}{2n+1}%
\right) ^{2n}W_{\phi }^{2}\text{,}\label{eebps}
\end{equation}
where we have used Eq. (\ref{ef1}) again.

For $n\geq 1$, the $\varepsilon _{BPS}$ above vanishes at $x=0$. So, the energy is expected to be distributed around the origin. Note that this is due to the nontrivial $f$. Moreover, once $W_\phi$ is supposed to vanish faster than $f$ when $x\rightarrow\pm\infty$, one must get $\varepsilon _{BPS} (x \rightarrow \pm \infty) \rightarrow 0$. This behavior means that the energy is localized, as desired. As a consequence, the total energy of the BPS solutions converges to the finite value given by Eq. (\ref{Ebps}) and does not depend on $f$.

In what follows, we go further and choose the superpotentials themselves. We then calculate the novel double-kink solutions analytically. We also address important aspects regarding the corresponding energy distributions and discuss some of the general properties.

%%%%%%%%%%%%%%%%%%%%%%%%%%%%%%%%%%%%%%%%%%%%%%%%%
{\subsection{The $\phi ^{4}$ case}}
%%%%%%%%%%%%%%%%%%%%%%%%%%%%%%%%%%%%%%%%%%%%%%%%%

We now explore a first example. We begin by considering the standard $\phi ^{4}$-superpotential. This superpotential was studied numerically in Ref. \cite{lca}. However, in that work, the authors adopted a different expression for the generalizing function. Even in this case, a comparison is possible, once the same general aspects are still present. Moreover, now we offer an analytical description of them, as we demonstrate below.

We choose the expression
\begin{equation}
W(\phi)=\phi-\frac{1}{3}\phi ^{3}\text{,}
\end{equation}%
from which one gets
\begin{equation}
W_{\phi }=1-\phi ^{2}\text{.}\label{dw}
\end{equation}%

In such a model, the kink is known to satisfy
\begin{equation}
\phi_\pm=\phi(x\rightarrow\pm\infty) \rightarrow \pm 1 \text{,} \label{ck4}
\end{equation}
which lead to $W\left( x\rightarrow +\infty \right)=-W\left( x\rightarrow -\infty
\right)=2/3$. The energy of the BPS kink can be then calculated as $E=E_{BPS}=4/3$, see Eq. (\ref{Ebps}). Furthermore, conditions (\ref{ck4}) guarantee that $W_{\phi }$ vanishes in the asymptotics, therefore confirming $\varepsilon _{BPS} (x \rightarrow \pm \infty) \rightarrow 0$.

In view of (\ref{dw}), Eq. (\ref{bebps}) can be written as
\begin{equation}
\frac{d\phi }{dy}=\pm \left( 1-\phi ^{2}\right) \text{,}
\end{equation}%
whose kink solution is%
\begin{equation}
\phi _{k}\left( y\right) =\tanh \left( y\right) \text{.} \label{ck40}
\end{equation}
We have assumed that the resulting profile is centered at $y=0$ (i.e. $x=0$), for the sake of illustration.

Now, given the relation (\ref{yx}), Eq. (\ref{ck40}) can be written explicitly in terms of $x$ as%
\begin{equation}
\phi _{k_{n}}\left( x\right) =\tanh \left( C_{n}x^{2n+1}\right) \text{,}\label{sphi4}
\end{equation}
which satisfies the conditions (\ref{ck4}), as expected for a $\phi^4$-kink.

%\bl{Thus, for the $\phi^4$-model, the generalizing function (\ref{ef1}) written in terms of the $\phi$-field reads
%\begin{equation}
%f\left( \phi \right) =\left[ \frac{1}{2}\ln \left( \frac{1+\phi }{1-\phi }%
%\right) \right] ^{-\frac{2n}{2n+1}}. \label{fgen_1}
%\end{equation}}

%%%%%%%%%%%%%%%%%%%%%%%%%%%%%%%%%%%%%%%%%%%%%%%%%%%%%%%%%%%%%%%%%%%%%%%%%%%
\begin{figure*}[!ht]
\begin{center}
  \centering
    \includegraphics[width=1.0 \textwidth]{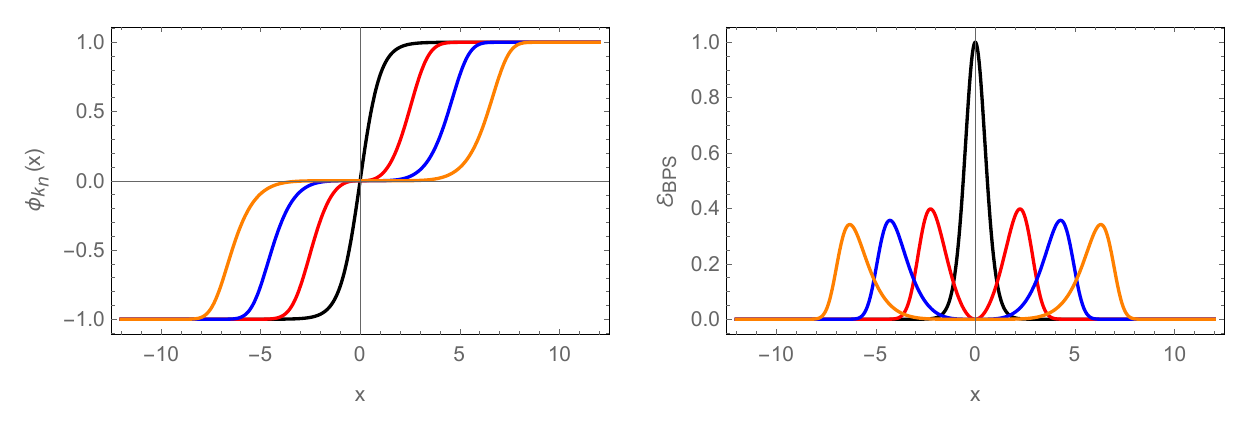}
    %\includegraphics[width=1.0 \textwidth]{F3d.pdf}\label{fig4a}
    % \subfigure[]{\includegraphics[width=0.42 \textwidth]{sGFig4b}\label{fig4b}}
    % \subfigure[]{\includegraphics[width=0.4115 \textwidth]{sGFig4c}\label{fig4c}}
    \vspace{-1.5cm}
  \caption{BPS double-kink solutions $\phi_{k_n}(x)$ (left) and their energy densities $\varepsilon _{BPS}$ (right) given by Eqs. (\ref{sphi4}) and (\ref{ephi4}), respectively. Here, $n=1$ (red line), $n=2$ (blue line), and $n=3$ (orange line). The black line represents $n=0$, i.e. the canonical single-kink solution.}
  \label{fig11xx}
\end{center}
\end{figure*}
%%%%%%%%%%%%%%%%%%%%%%%%%%%%%%%%%%%%%%%%%%%%%%%%%%%%%%%%%%%%%%%%%%%%%%%%%%

We depict the BPS solution (\ref{sphi4}) explicitly in Fig. \ref{fig11xx} (left) for different values of $n$. The double-kink profiles attained for $n \geq 1$ are now evident. As mentioned previously, these analytical solutions develop a plateau around $x=0$. In this sense, they mimic the purely numerical structure found in \cite{lca}. However, it is now explained on an exact basis. Asymptotically, the novel solutions converge to the vacuum states predicted by the $\phi^4$-superpotential; the reader may confront our results with Fig. 11(a) of Ref. \cite{lca}.

To confirm the legitimate double-kink behavior, we study how $\phi _{k_{n}}$ itself approaches the origin. In this limit, Eq. (\ref{sphi4}) reduces to
\begin{equation}
\phi _{k_{n}}\left( x\rightarrow 0\right) \approx C_{n}x^{2n+1}\text{,}\label{phi4no}
\end{equation}%
which proves that we have obtained a genuine double-kink solution. It is interesting to note that (\ref{phi4no}) reproduces exactly Eq. (48) (with an upper sign) of Ref. \cite{lca} for $m=\lambda=v=1$ and $z=n$.

We also consider how the generalizing function affects the way the $\phi^4$-field reaches its vacuum values. In the asymptotic regions, Eq. (\ref{sphi4}) behaves as
\begin{equation}
\phi _{k_{n}}\left( x\rightarrow \pm \infty \right) \approx \pm 1\mp
2e^{\mp 2C_{n}x^{2n+1}}\text{.}\label{phi4exp}
\end{equation}%
That is, the typical exponential decay remains preserved even in the presence of a nontrivial $f$. However, the generalizing function controls such a decay by determining not only its dependence on $x$ (through the power $2n+1$), but also the mass of the BPS double-kink (through $C_{n}$). For the sake of comparison, we point out that the numerical profiles introduced in Ref. \cite{lca} also possess an exponential tail. In that case, however, the generalizing function controls only the mass of the double-kinks, see Eqs. (47), (49) and (50) of that work. This is due to the different expression that those authors adopted for $f$.

Now, in view of Eqs. (\ref{dw}) and (\ref{sphi4}), the BPS energy distribution (\ref{eebps}) can be expressed in the form
\begin{equation}
\varepsilon _{BPS}=\left( \frac{x}{2n+1}\right) ^{2n}\text{sech}^{4}\left(
C_{n}x^{2n+1}\right) \text{,} \label{ephi4}
\end{equation}
whose profiles appear in Fig. \ref{fig11xx} (right). Here, the $n=0$ solution is the standard single-lump one. On the other hand, the two-lump profiles emerge  for $n \geq 1$. Naturally, each lump refers to an individual kink in the double-kink configuration. Note that two-lump profiles also appear in Fig. 12 of \cite{lca}.

These lumps are placed symmetrically with respect to $x=0$. However, when taken individually, they are not symmetric with respect to their own center. This is due to the fact that their inner tails refer to the power-like approximation (\ref{phi4no}), while the outer ones reflect the exponential decay (\ref{phi4exp}). The numerical solutions explored in \cite{lca} exhibit the same aspect.

Moreover, Fig. \ref{fig11xx} (right) indicates that $n$ controls not only the positions of the individual lumps, but also their heights. In general, as $n \neq 0$ increases, these lumps move away from the origin, while their heights decrease. In this regard, Eq. (\ref{ephi4}) leads to
\begin{equation}
\varepsilon' _{BPS}(x)=\frac{4(2n+1)}{x}\left[\frac{n}{2(2n+1)} - C_nx^{2n+1} \tanh{\left(C_nx^{2n+1}\right)} \right] \varepsilon _{BPS}(x) \text{,}
\end{equation}
which reveals that the individual lumps have their peaks positioned at $x= x_n$ such that
\begin{equation}
C_nx_n^{2n+1} \tanh{\left(C_n x_n^{2n+1}\right)}=\frac{n}{2(2n+1)} \text{.} \label{te}
\end{equation}

Equation (\ref{te}) is a transcendental one whose roots must be determined numerically for each $n$. Naturally, $n=0$ leads to $x_0=0$ (i.e. the standard single-lump is centered at the origin). In addition, for $n=1$, $n=2$, and $n=3$, one gets $x_1 \approx \pm 2.24656$, $x_2 \approx \pm 4.28574$, and $x_3 \approx \pm 6.30340$, respectively.

In addition, these roots allow us to calculate the peaks themselves. In this way, we get $\varepsilon _{BPS}(x_0) = 1 $ (the usual $n=0$ result), $\varepsilon _{BPS}(x_1) \approx 0.398031$ (for $n=1$), $\varepsilon _{BPS}(x_2) \approx 0.356913$ ($n=2$), and $\varepsilon _{BPS}(x_3) \approx 0.3419$ ($n=3$), see Eq. (\ref{ephi4}) itself.

All these values agree very well with the results pointed out in Fig. \ref{fig11xx}.

\vspace{0.4cm}

%%%%%%%%%%%%%%%%%%%%%%%%%%%%%%%%%%%%%%%%%%%%%%%%%
{\subsection{The $\phi^{6}$ case}}
%%%%%%%%%%%%%%%%%%%%%%%%%%%%%%%%%%%%%%%%%%%%%%%%%

We now consider a second example. It is defined by the $\phi ^{6}$-superpotential. This model was not studied in Ref. \cite{lca}. So, a direct comparison is not possible. Even in this case, the basic aspects are still present, as we demonstrate.

We choose
\begin{equation}
W(\phi)= \frac{1}{2}\phi^2-\frac{1}4{}\phi ^{4} \text{,}
\end{equation}%
from which we get
\begin{equation}
W_{\phi }=\phi \left( 1-\phi ^{2}\right) \text{.} \label{dw0}
\end{equation}%

The corresponding kink then satisfies
\begin{equation}
\phi _{-}\left( x\rightarrow -\infty \right) \rightarrow 0 \text{ \ \ and \ \ }\phi _{+}\left( x\rightarrow
+\infty \right) \rightarrow +1\text{.}\label{ck6}
\end{equation}%
These values lead to $W\left( x\rightarrow +\infty \right)=1/4$, while $W\left( x\rightarrow -\infty \right)$ vanishes. The energy of the BPS solutions is then $E=E_{BPS}=1/4$, see Eq. (\ref{Ebps}). These conditions also guarantee that the corresponding energy distribution is localized.

In this case, Eq. (\ref{bebps}) assumes the form
\begin{equation}
\frac{d\phi }{dy}=\pm \phi \left( 1-\phi ^{2}\right) \text{,}
\end{equation}%
whose kink solution can be promptly verified to be%
\begin{equation}
\phi _{k}\left( y\right) =\frac{1}{\sqrt{1+e^{-2y}}}\text{,}\label{sphi60}
\end{equation}
which was assumed to be positioned at the origin, for simplicity.

In view of (\ref{yx}), we rewrite Eq. (\ref{sphi60}) as
\begin{equation}
\phi _{k_{n}}\left( x\right) =\frac{1}{\sqrt{1+e^{-2C_{n} x^{2n+1}}}}\text{,}\label{sphi6}
\end{equation}
which behaves as the conditions (\ref{ck6}), as expected.

%\bl{As expected, we now can write the generalizing function (\ref{ef1}) written in terms of the $\phi$-field as
%\begin{equation}
%f\left( \phi \right) =\left[ \frac{1}{2}\ln \left( \frac{\phi ^{2}}{1-\phi
%^{2}}\right) \right] ^{-\frac{2n}{2n+1}}. \label{fgen_2}
%\end{equation}}

%%%%%%%%%%%%%%%%%%%%%%%%%%%%%%%%%%%%%%%%%%%%%%%%%%%%%%%%%%%%%%%%%%%%%%%%%%%
\begin{figure*}[!ht]
\begin{center}
  \centering
    \includegraphics[width=1.0 \textwidth]{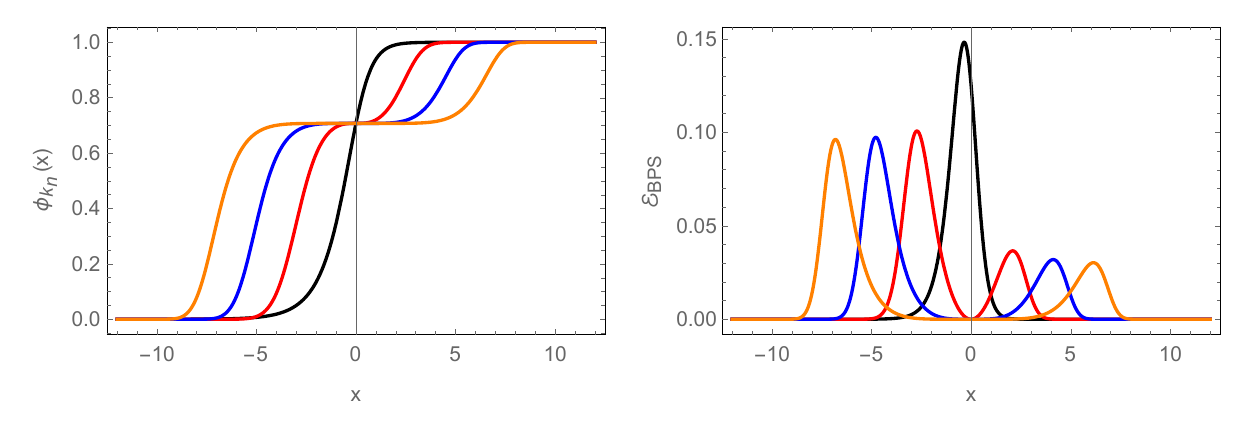}
    %\includegraphics[width=1.0 \textwidth]{F3d.pdf}\label{fig4a}
    % \subfigure[]{\includegraphics[width=0.42 \textwidth]{sGFig4b}\label{fig4b}}
    % \subfigure[]{\includegraphics[width=0.4115 \textwidth]{sGFig4c}\label{fig4c}}
    \vspace{-1.5cm}
  \caption{BPS solutions $\phi_{k_n}(x)$ (left) and their energy densities $\varepsilon _{BPS}$ (right) given by Eqs. (\ref{sphi6}) and (\ref{ephi6}), respectively. Conventions as in Fig. \ref{fig11xx}.}
  \label{fig22xx1}
\end{center}
\end{figure*}
%%%%%%%%%%%%%%%%%%%%%%%%%%%%%%%%%%%%%%%%%%%%%%%%%%%%%%%%%%%%%%%%%%%%%%%%%%

Figure \ref{fig22xx1} (left) shows the solutions (\ref{sphi6}) for different $n$. As before, double-kink profiles emerge for $n \geq 1$. They again exhibit a plateau around the origin, while converging to the states inherent to the $\phi^6$-model. The usual solution is again depicted, for comparison.

It is interesting to note that the double-kink solutions preserve certain {\it{asymmetry}} between the individual kinks. That is, the kink on the left of $x=0$ is higher than the one on the right. This asymmetry stands for a novel effect once it was not explored in Ref. \cite{lca}. Naturally, it also influences the corresponding energy distribution. We return to this point below.

Near the origin, Eq. (\ref{sphi6}) behaves as
\begin{equation}
\phi _{k_{n}}\left( x\rightarrow 0\right) \approx \frac{1}{\sqrt{2}}+%
\frac{C_{n}\sqrt{2}}{4}x^{2n+1}\text{,}\label{haha}
\end{equation}%
i.e. it reveals that (\ref{sphi6}) itself represents a legitimate kink solution. In addition, note that Eqs. (\ref{haha}) and (\ref{phi4no}) depend on $x$ in the same way. So, it is reasonable to infer that such a dependence is due to the generalizing function itself.

On the other hand, near the asymptotic boundaries, $\phi_{k_n}$ can be approximated by
\begin{equation}
\phi _{k_{n}}\left( x\rightarrow
-\infty \right) \approx e^{C_{n}x^{2n+1}}\text{ \ \ and \ \ }\phi _{k_{n}}\left( x\rightarrow +\infty \right) \approx 1-\frac{1}{2}%
e^{-2C_{n}x^{2n+1}}\text{,}
\end{equation}
which indicates that the exponential decay remains preserved even for $n \neq 0$. The generalizing function controls this decay in the same way as before, i.e. via its dependence on $x$ and the mass of the kink.

The energy density (\ref{eebps}) can be written as
\begin{equation}
\varepsilon _{BPS}=\left( \frac{x}{2n+1}\right) ^{2n}\frac{%
e^{-4C_{n}x^{2n+1}}}{\left( 1+e^{-2C_{n}x^{2n+1}}\right)^{3}}\text{,} \label{ephi6}
\end{equation}
where we have used Eqs. (\ref{dw0}) and (\ref{sphi6}). The corresponding solutions are shown in Fig. \ref{fig22xx1} (right). The single-lump profile is the usual result ($n=0$), while two-lump configurations arise when $n \neq 0$. Again, each one of these lumps refers to an individual kink in the double-kink solution.

The individual lumps are not symmetric with respect to their own center, and such an asymmetry can be explained in the same way as before. Moreover, unlike the previous example, those lumps with the same $n \neq 0$ are {\it{not}} positioned symmetrically with respect to the origin. Their heights are also different, with the lump on the left being higher than that on the right. Such an asymmetric energy distribution was not explored in Ref. \cite{lca}.

These differences are related to the asymmetry between the individual kinks that appear from Eq. (\ref{sphi6}) for a fixed $n \neq 0$. So, in order to explain this asymmetry on a quantitative basis, we now calculate both the position and the height of each lump explicitly. First, we write
\begin{equation}
\varepsilon' _{BPS}=\frac{2(2n+1)}{x}\left[\frac{n}{2n+1} - \left( \tanh \left( C_nx^{2n+1} \right) + \frac{e^{C_nx^{2n+1}}}{2} \sech \left( C_nx^{2n+1} \right) \right) C_nx^{2n+1}  \right] \varepsilon _{BPS} \text{,}
\end{equation}
%\begin{equation}
%\varepsilon' _{BPS}(x)=\frac{2(2n+1)}{x}\left(\frac{n}{2n+1} - \left( \right) \frac{2-e^{-2C_nx^{2n+1}}}{1+e^{-2C_nx^{2n+1}}}C_nx^{2n+1}  \right) \varepsilon _{BPS}(x) \text{,}
%\end{equation}
which states that the position $x=x_n$ of each individual lump satisfies
\begin{equation}
\left[ \tanh \left( C_nx_n^{2n+1} \right) + \frac{e^{C_nx_n^{2n+1}}}{2} \sech \left( C_nx_n^{2n+1} \right) \right] C_nx_n^{2n+1}=\frac{n}{2n+1} \text{.} \label{fe00x}
\end{equation}
%\begin{equation}
%\frac{2-e^{-2C_nx_n^{2n+1}}}{1+e^{-2C_nx_n^{2n+1}}}C_nx_n^{2n+1}=\frac{n}{2n+1} \text{.} \label{fe00x}
%\end{equation}
Again, these roots must be calculated numerically for each $n$ fixed.

For $n=0$, Eq. (\ref{fe00x}) provides $x_0 \approx -0.346574$, i.e. the location of the usual single-lump solution. Moreover, as mentioned previously, those lumps for $n \geq 1$ are not symmetrically placed with respect to the origin. So, we define $x_{l_n}$ ($x_{r_n}$) as the position of the lump that lies on the left (right) of $x=0$. Then, for $n=1$, $n=2$, and $n=3$, Eq. (\ref{fe00x}) gives rise to $x_{l_1} \approx -2.71896$ and $x_{r_1} \approx 2.08861$, $x_{l_2} \approx -4.78499$ and $x_{r_2} \approx 4.12485$, and $x_{l_3} \approx -6.81255$ and $x_{r_3} \approx 6.14198$, respectively. Therefore, the position values effectively verify that the corresponding lumps, for a given $n\geq 1$, are not symmetrically located.

At these positions, the peaks can be calculated as $\varepsilon _{BPS}(x_0) \approx 0.148148$ (the canonical result), $\varepsilon _{BPS}(x_{l_1}) \approx 0.100662$ and $\varepsilon _{BPS}(x_{r_1}) \approx 0.0365608$, $\varepsilon _{BPS}(x_{l_2}) \approx 0.0972757$ and $\varepsilon _{BPS}(x_{r_2}) \approx 0.0319025$, and $\varepsilon _{BPS}(x_{l_3}) \approx 0.096144$ and $\varepsilon _{BPS}(x_{r_3}) \approx 0.0302369$, see Eq. (\ref{ephi6}).

The values above illustrate the asymmetry on a quantitative basis. In this sense, they explain the dimensions of the profiles in Fig. \ref{fig22xx1}. We reinforce that such an asymmetric energy distribution is an original result that was not found in Ref. \cite{lca}.

\vspace{0.4cm}

%%%%%%%%%%%%%%%%%%%%%%%%%%%%%%%%%%%%%%%%%%%%%%%%%
{\subsection{The sine-Gordon case}}
%%%%%%%%%%%%%%%%%%%%%%%%%%%%%%%%%%%%%%%%%%%%%%%%%

We end this Section by briefly investigating the sine-Gordon field. This model is particularly interesting due to the integrability of its canonical version. In such a case, the kink-antikink collision does not allow for the escape of energy radiation, for instance.

The sine-Gordon superpotential is
\begin{equation}
W(\phi)= - \cos \phi \text{,}
\end{equation}%
whose derivative reads
\begin{equation}
W_{\phi }=\sin \phi \text{.}\label{dwsg}
\end{equation}

The sine-Gordon kink is known to behave as
\begin{equation}
\phi _{-}\left( x\rightarrow -\infty \right) \rightarrow 0 \text{ \ \ and \ \ }\phi _{+}\left( x\rightarrow
+\infty \right) \rightarrow +\pi\text{,}\label{cksg}
\end{equation}%
via which we calculate $W\left( x\rightarrow \pm \infty \right)= \pm 1$. The BPS total energy can then be verified to be $E=E_{BPS}=2$, while $\varepsilon _{BPS} (x \rightarrow \pm \infty) \rightarrow 0$ is promptly satisfied.

In the present case, Eq. (\ref{bebps}) can be written as
\begin{equation}
\frac{d\phi }{dy}=\pm \sin \phi \text{,}
\end{equation}%
whose kink solution (centered at $y=0$) reads
\begin{equation}
\phi _{k}\left( y\right) =2\arctan \left( e^{y}\right) \text{.}
\end{equation}

\begin{figure*}[!ht]
\begin{center}
  \centering
    \includegraphics[width=1.0 \textwidth]{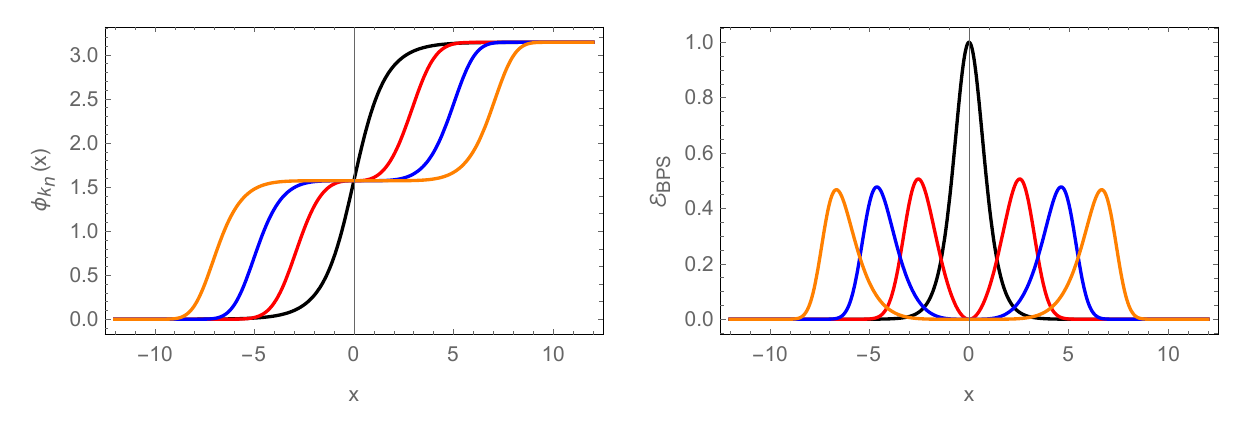}\label{fig22xx}
    %\includegraphics[width=1.0 \textwidth]{F3d.pdf}\label{fig4a}
    % \subfigure[]{\includegraphics[width=0.42 \textwidth]{sGFig4b}\label{fig4b}}
    % \subfigure[]{\includegraphics[width=0.4115 \textwidth]{sGFig4c}\label{fig4c}}
    \vspace{-1.5cm}
  \caption{BPS solutions $\phi_{k_n}(x)$ (left) and their energy densities $\varepsilon _{BPS}$ (right) given by Eqs. (\ref{sphisg}) and (\ref{ephisg}), respectively. Conventions as in Fig. \ref{fig11xx}.}
  \label{fig33xx}
\end{center}
\end{figure*}

Then, in view of (\ref{yx}), we obtain
\begin{equation}
\phi _{k_{n}}\left( x\right) =2\arctan \left( e^{C_{n}x^{2n+1}}\right) \text{,}\label{sphisg}
\end{equation}
%\bl{which allows us to express the generalizing function (\ref{ef1}) in terms of the $\phi$-field as
%\begin{equation}
%f\left( x\right) =\left[ \ln \left( \tan \frac{\phi }{2}\right) \right] ^{-%
%\frac{2n}{2n+1}}. \label{fgen_3}
%\end{equation}}
whose profiles appear in Fig. \ref{fig33xx} (left) for different $n$. As in the $\phi^4$ case, the individual kinks in the same double-kink solution are symmetric with respect to each other. That is, they have the same height and width. We highlight that all solutions satisfy the conditions (\ref{cksg}).

Near the origin and asymptotically, $\phi_{k_n}$ can be written as%
\begin{equation}
\varphi _{k_{n}}\left( x\rightarrow 0\right) \approx \frac{\pi }{2}%
+C_{n} x^{2n+1}
\end{equation}%
and
\begin{equation}
\varphi _{k_{n}}\left( x\rightarrow +\infty \right) \approx \pi
-2e^{-C_{n}x^{2n+1}}\text{ \ \ and \ \ }\varphi _{k_{n}}\left( x\rightarrow
-\infty \right) \approx 2e^{C_{n}x^{2n+1}}\text{,}
\end{equation}
respectively. These expressions confirm that the genuine kink preserves the exponential tail for any $n \ge 0$ integer. The dependence on $x$ is the same as in the previous cases, and the generalizing function again controls the mass of the field.

%%%%%%%%%%%%%%%%%%%%%%%%%%%%%%%%%%%%%%%%%%%%%%%%%%%%%%%%%%%%%%%%%%%%%%%%%%%
%%%%%%%%%%%%%%%%%%%%%%%%%%%%%%%%%%%%%%%%%%%%%%%%%%%%%%%%%%%%%%%%%%%%%%%%%%

Figure \ref{fig33xx} (right) shows the BPS energy distribution. In this case, the analytical solution reads%
\begin{equation}
\varepsilon _{BPS}=\left( \frac{x}{2n+1}\right) ^{2n}\frac{4e^{2C_{n}x^{2n+1}}}{\left( 1+e^{2C_{n} x^{2n+1}}\right) ^{2}}\text{,}\label{ephisg}
\end{equation}
where we have used (\ref{dwsg}) and (\ref{sphisg}). The standard solution ($n=0$) stands for a single-lump centered at $x=0$. For $n \geq 1$, two-lump configurations appear. As in the previous $\phi^4$ case, the individual lumps with the same $n$ are symmetrically placed w.r.t. the origin. Also, their peaks reach the same height.

To confirm this symmetry, we calculate both the position and height of the individual lumps. We write
\begin{equation}
\varepsilon' _{BPS}(x)=\frac{2(2n+1)}{x}\left[\frac{n}{2n+1} - C_nx^{2n+1} \tanh \left( C_nx^{2n+1} \right)  \right] \varepsilon _{BPS}(x) \text{,}
\end{equation}
which reveals that the positions of the individual lumps are given as the roots of
\begin{equation}
C_nx_n^{2n+1}\tanh \left( C_nx_n^{2n+1} \right) = \frac{n}{2n+1} \text{.}\label{tesg}
\end{equation}
Note that, except for a multiplicative factor, it is similar to Eq. (\ref{te}), whose solutions define the positions of the individual lumps in the $\phi^4$ case.

By solving Eq. (\ref{tesg}) numerically for fixed values of $n$, we find that the individual lumps are placed at $x= x_n$, such that $x_0=0$ (usual result), $x_1 \approx \pm 2.5463$ ($n=1$), $x_2 \approx \pm 4.62589$ ($n=2$), and $x_3 \approx \pm 6.65936$ ($n=3$). At these points, the peaks are $\varepsilon _{BPS}(x_0) = 1$, $\varepsilon _{BPS}(x_1) \approx 0.506308$, $\varepsilon _{BPS}(x_2) \approx 0.477522$, and $\varepsilon _{BPS}(x_3) \approx 0.467563$, see Eq. (\ref{ephisg}).

\vspace{0.5cm}
%%%%%%%%%%%%%%%%%%%%%%%%%%%%%%%%%%%%%%%%%%%%%%%%%
\section{Summary and perspectives} \label{secIV}
%%%%%%%%%%%%%%%%%%%%%%%%%%%%%%%%%%%%%%%%%%%%%%%%%

We have found exact analytic BPS solutions with a double-kink profile by studying a $(1+1)$-dimensional model that contains a single real scalar field $\phi$. These novel and original solutions have allowed us to explain essential aspects of BPS double-kinks quantitatively. To generate these new configurations, we have enlarged the kinematics of the scalar field to accommodate a generalizing function $f(\phi)$. In such circumstances, the BPS potential $V(\phi)$ is required to depend on $f$ itself, and, as a consequence, the corresponding BPS equation also contains this function explicitly; see (\ref{EEbps}).

To solve the BPS Eq. (\ref{EEbps}), we have introduced the new coordinate $y$. It is related to the original coordinate $x$ through the generalizing function $f(\phi(x))\equiv f(x)$. So, we have assumed that it is allowed to propose $f$ as a function of $x$ instead of $\phi$, see Eq. (\ref{ef1}). As a consequence, we have obtained an analytical relation between $y$ and $x$, see Eq. (\ref{yx}). Here, we have supposed that Eq. (\ref{bebps}) provides an analytical solution $\phi(y(x))$ that can be inverted to provide $x(\phi)$. %One has verified our proposal through the three models studied here, as shown by the results given in Eqs. (\ref{fgen_1}), (\ref{fgen_2}), and (\ref{fgen_3}).

We have focused on those superpotentials that define the standard $\phi^4$, $\phi^6$, and sine-Gordon models, separately. In all cases, we have obtained exact BPS double-kinks, see Eqs. (\ref{sphi4}), (\ref{sphi6}), and (\ref{sphisg}), respectively. We have depicted the novel profiles on the left side of Figs. \ref{fig11xx}, \ref{fig22xx1}, and \ref{fig33xx}. For the sake of elucidation, we have also compared them to the numerical ones studied in \cite{lca}, from which we have pointed out similarities and differences between the two scenarios.

Moreover, we have written analytical expressions of the energy density of the new configurations, see Eqs. (\ref{ephi4}), (\ref{ephi6}), and (\ref{ephisg}). In addition, we have depicted them on the right side of the Figs. \ref{fig11xx}, \ref{fig22xx1}, and \ref{fig33xx}, respectively. We have then observed that these densities assume a two-lump profile. In particular, the individual lumps may not necessarily be symmetric in relation to the origin. Therefore, to explain this aspect quantitatively, we have calculated both the position and the peak (maximum value) of all individual lumps.

Because the double-kink solutions presented here allow us to express analytically $x\equiv x(\phi)$, it is possible to represent both the BPS potential (\ref{ee5}) and the BPS energy density (\ref{edx}) as explicit functions of $\phi$. Since both quantities are proportional (i.e. $\varepsilon _{BPS}=2V$), it becomes clear that the BPS potential also converges to a vacuum state in the asymptotic limits, as established in (\ref{cp}). Similarly, our main hypothesis (i.e. that $f$ can be treated directly as a function $x$) is effectively satisfied.

It is also important to highlight that $f(\phi)$ cannot be wholly absorbed into $\phi$ via a field redefinition, as discussed in Appendix A of Ref. \cite{lca}. This way, the double-kinks raised in this manuscript are genuine new BPS configurations, very different from the standard single-kink ones. In other words, the latest solutions encountered here are not a mere redefinition of the canonical results. Finally, the approach developed in this work can be applied to investigate novel configurations in connection with other superpotentials.

Perspectives regarding future developments include the study of double-kink-double-antikink collisions. In a sine-Gordon scenario, such a collision may give rise to interesting phenomena, such as the formation of a bound state with two oscillons, or may also lead to a novel resonance pattern. In the last case, the determination of the linear excitation spectrum and its application to describe the corresponding resonance phenomena is also an issue of interest. We are currently working on these topics, and positive results will be presented in a forthcoming manuscript.

%%%%%%%%%%%%%%%%%%%%%%%%%%%%%%%%%%%%%%%%%%%%%%

\section*{Acknowledgements}

We thank Fundação de Amparo à Pesquisa e ao Desenvolvimento Científico e Tecnológico do Maranhão (FAPEMA), Conselho Nacional de Desenvolvimento Científico e Tecnológico (CNPq), and Coordenação de Aperfeiçoamento de Pessoal de Nível Superior (CAPES) - Finance Code 001 (Brazilian agencies) for partial financial support. R. C. acknowledges the support from the grants CNPq/312155/2023-9, FAPEMA/UNIVERSAL 00812/19, and FAPEMA APP-12299/22.

%%%%%%%%%%%%%%%%%%%%%%%%%%%%%%%%%%%%%%%%%%%%%%

%%%%%%%%%%%%%%%%%%%%%%%%%%%%%%%%%%%%%%%%%%%%%%


\begin{thebibliography}{99}

%%%%%%%%%%%%%%%%%%%%%%%%%%%%%%%%%%%%%%%%%%%%%%

\bibitem{masu} N. Manton and P. Sutcliffe, {\it{Topological Solitons}}, Cambridge University Press, (2004).

\bibitem{Enz}U. Enz, {\it{Discrete Mass, Elementary Length, and a Topological Invariant as a Consequence of a Relativistic Invariant Variational Principle}}, Phys. Rev. {\bf 131}, 1392 (1963).

\bibitem{Fin}D. Finkelstein, {\it{Kinks}}, J. Math. Phys. {\bf 7}, 1218 (1966).

\bibitem{DHN}R. F. Dashen, B. Hasslacher and A. Neveu, {\it{Nonperturbative methods and extended-hadron models in field theory. II. Two-dimensional models and extended hadrons}}, Phys. Rev. D {\bf 10}, 4130 (1974).

\bibitem{prasom} M. K. Prasad and Charles M. Sommerfield, {\it Exact Classical Solution for the 't Hooft Monopole and the Julia-Zee Dyon}, {{Phy. Rev. Lett.} {\bf 35}, 12 (1975)}.

\bibitem{bogo} E. B. Bogomolny, {\it Stability of Classical Solutions}, {{Sov. J. Nucl. Phys.} {\bf 24}, 449 (1976)}.

\bibitem{Gri}D. Yu. Grigoriev and V. A. Rubakov, {\it{Soliton pair creation at finite temperatures: Numerical study in (1+1) dimensions}}, Nucl. Phys. B {\bf 299}, 67 (1988).

\bibitem{Man}N. Manton and T. M. Samols, {\it{Sphalerons on a circle}}, Phys. Lett. B
{\bf 207}, 179 (1988).

\bibitem{Vacha2} N. D. Antunes, L. Pogosian and T. Vachaspati, {\it{Formation of domain wall lattices}}, Phys. Rev. D {\bf 69}, 043513 (2004).

\bibitem{Vacha1} L. Pogosian and T. Vachaspati, {\it{Domain wall lattices}}, Phys. Rev. D {\bf 67}, 065012 (2003).

\bibitem{Man2}N. Manton, {\it{Integration theory for kinks and sphalerons
in one dimension}}, J. Phys. A {\bf 57},
025202 (2024).

\bibitem{blm} D. Bazeia, M. A. Liao and M. A. Marques, {\it{Geometrically constrained kinklike configurations}}, Eur. Phys. J. Plus {\bf{135}}, 383 (2020).

\bibitem{balbazmar} A. J. Balseyro, D. Bazeia and M. A. Marques, {\it{Mechanism to induce geometric constriction on kinks and domain walls}}, Eur. Phys. Lett. {\bf{141}}, 34003 (2023).

\bibitem{bazmarmen} D. Bazeia, M. A. Marques and R. Menezes, {\it{Geometrically constrained kink-like configurations engendering long-range, double-exponential, half-compact and compact behavior}}, Eur. Phys. J. Plus {\bf{138}}, 735 (2023).

\bibitem{jfd} João G. F. Campos, Fabiano C. Simas and D. Bazeia, {\it Kink scattering in the presence of geometric constrictions}, J. High Energy Phys. {\bf{10}}, 124 (2023).

\bibitem{marmen} M. A. Marques and R. Menezes, {\it{Geometrically constrained multifield models with BNRT solutions}}, Chaos, Solitons and Fractals {\bf{181}}, 114730 (2024).

\bibitem{bs} D. Bazeia and G. S. Santiago, {\it{Kink crystal}}, Eur. Phys. J. C {\bf{84}}, 323 (2024).

\bibitem{diosimas} D. Bazeia and F. C. Simas, {\it Fermion bound states from Yukawa coupling with periodic bosonic background}, Eur. Phys. J. C {\bf{84}}, 1039 (2024).

\bibitem{hora} E. da Hora, L. Pereira, C. dos Santos and F. C. Simas, {\it Geometrically constrained sine-Gordon field: BPS solitons and their collisions}, Commun. Nonlinear Sci. Numer. Simul. {\bf{151}}, 109070 (2025).

\bibitem{hora1} E. da Hora, C. dos Santos and Fabiano C. Simas, {\it Sine-Gordon kink lattice}, Eur. Phys. J. C {\bf{85}}, 481 (2025).

\bibitem{poga} S. D. Pollard et al., {\it{Bloch chirality induced by an interlayer Dzyaloshinskii-Moriya interaction in ferromagnetic multilayers}}, Phys. Rev. Lett. {\bf{125}}, 227203 (2020).

\bibitem{chdo} J. Chen and S. Dong, {\it{Manipulation of magnetic domain walls by ferroelectric switching: dynamic magnetoelectricity at the nanoscale}}, Phys. Rev. Lett. {\bf{126}}, 117603 (2021).

\bibitem{Rev}M. Buballa and S. Carignano, {\it{Inhomogeneous chiral condensates}}, Prog. Part. Nucl. Phys. {\bf 81}, 39 (2015).

\bibitem{Mat1}M. Matsumoto, S. Nakamura and R. Yoshii, {\it{Kink crystalline condensate and multi-kink solution in holographic superconductor}}, J. High Energy Phys. {\bf 2020}, 22 (2020).

\bibitem{Mat2}M. Matsumoto and R. Yoshii, {\it{Twisted kink crystal in holographic superconductor}}, Phys. Rev. D
{\bf 104}, 066007 (2021).

\bibitem{Dun1}G. Basar and G. V. Dunne, {\it{Self-Consistent Crystalline Condensate in Chiral Gross-Neveu and Bogoliubov–de Gennes Systems}}, Phys. Rev. Lett.
{\bf 100}, 200404 (2008).

\bibitem{Dun2}G. Basar and G. V. Dunne, {\it{Twisted kink crystal in the chiral Gross-Neveu model}}, Phys. Rev. D
{\bf 78}, 065022 (2008).

\bibitem{Dun3}G. Basar, G. V. Dunne and M. Thies, {\it{Inhomogeneous condensates in the thermodynamics of the chiral NJL$_2$ model}}, Phys. Rev. D {\bf 79}, 105012 (2009).

\bibitem{c10}Y. Zhong, X. -L. Du, Z. -C. Jiang, Y. -X. Liu and Y. -Q. Wang, {\it{Collision of two kinks with inner structure}}, J. High Energy Phys. {\bf 02}, 153 (2020).

\bibitem{c11}E. Belendryasova and V. A. Gani, {\it{Scattering of the $\phi^8$ kinks with power-law asymptotics}}, Commun. Nonlinear Sci. Numer. Simul. {\bf 67}, 414 (2019).

\bibitem{c12}B. A. Mello, J. A. Gonzalez, L. E. Guerrero and E. Lopez-Atencio, {\it{Topological defects with long-range interactions}}, Phys. Lett. A {\bf 244}, 277 (1998).

\bibitem{c13}A. Khare, I. C. Christov and A. Saxena, {\it{Successive phase transitions and kink solutions in $\phi^8$, $\phi^{10}$, and $\phi^{12}$ field theories}}, Phys. Rev. E {\bf 90}, 023208 (2014).

\bibitem{c14}D. Saadatmand, S. V. Dmitriev and P. G. Kevrekidis, {\it{High energy density in multisoliton collisions}}, Phys. Rev. D {\bf 92}, 056005 (2015).

\bibitem{c15}A. M. Marjaneh, D. Saadatmand, K. Zhou, S. V. Dmitriev and M. E. Zomorrodian, {\it{High energy density in the collision of N kinks in the $\phi^4$ model}}, Commun. Nonlinear Sci. Numer. Simul. {\bf 49}, 30 (2017).

\bibitem{c16}V. A. Gani, A. M. Marjaneh and D. Saadatmand, {\it{Multi-kink scattering in the double sine-Gordon model}}, Eur. Phys. J. C {\bf 79}, 620 (2019).

\bibitem{c17}L. Losano, M. A. Marques and R. Menezes, {\it{Generalized scalar field models with the same energy density and linear stability}}, Phys. Lett. B {\bf 775}, 178 (2017).

\bibitem{lca} F. C. E. Lima, R. Casana and C. A. S. Almeida, {\it{Kinks and double-kinks in generalized $\phi^4$- and $\phi^8$-models}}, Eur. Phys. J. C {\bf 84}, 1266 (2024).

\end{thebibliography}
\end{document}